\documentclass[onecolumn]{IEEEtran}
\usepackage{multirow}
\usepackage{wrapfig}
\usepackage{float}
\usepackage{tabularx}
\usepackage{longtable}
\usepackage{color}
\definecolor{menucolor}{rgb}{0.1,0.52,0.47}
\definecolor{anchorcolor}{rgb}{0.85,0.37,0.01}
\definecolor{runcolor}{rgb}{0.46,0.44,0.701}
\definecolor{linkcolor}{rgb}{0.3,0.55,0.01}
\definecolor{urlcolor}{rgb}{0.12,0.47,0.70}
\definecolor{citecolor}{rgb}{0.55,0.36,0.01}
\definecolor{filecolor}{rgb}{0.4,0.4,0.4}
\usepackage[colorlinks=true, urlcolor=urlcolor, linkcolor=linkcolor,citecolor=citecolor,filecolor=filecolor, anchorcolor=anchorcolor,menucolor=menucolor]{hyperref}
\usepackage{graphicx}
\usepackage{verbatim}
\usepackage{booktabs}
\usepackage{tikz}
\ifCLASSOPTIONcompsoc
  \usepackage[caption=false,font=normalsize,labelfont=sf,textfont=sf]{subfig}
\else
\usepackage[caption=false,font=footnotesize]{subfig}
\fi
\usepackage{array}

\usepackage{multirow}
\usepackage{wrapfig}
\usepackage{tabularx}
\usepackage{longtable}
\usepackage{hyperref}
\usepackage{amsmath}
\usepackage{amsthm}
\usepackage{graphicx}
\graphicspath{{./figures/}}
\usepackage{verbatim}
\usepackage[nocompress]{cite}

\usepackage{xcolor}
\usepackage{tikz}
\usetikzlibrary{shapes.geometric}
\definecolor{winep1}{RGB}{27,   158,  119}
\definecolor{winep2}{RGB}{217, 95,    2}
\definecolor{winep3}{RGB}{117,   112,  179}

\definecolor{oilp1}{RGB}{27,   158,  119}
\definecolor{oilp2}{RGB}{217, 95,    2}
\definecolor{oilp3}{RGB}{117,   112,  179}
\definecolor{oilp4}{RGB}{231,    41,  138}
\definecolor{oilp5}{RGB}{102,   166,   30}
\definecolor{oilp6}{RGB}{230,   171,  2}
\definecolor{oilp7}{RGB}{166,   118,  29}
\definecolor{oilp8}{RGB}{102,   102,  102}
\definecolor{oilp9}{RGB}{0,    255,   255}

\definecolor{cbp1}{RGB}{166,206,227}
\definecolor{cbp2}{RGB}{31,120,180}
\definecolor{cbp3}{RGB}{253,191,111}
\definecolor{cbp4}{RGB}{255,127,0}
\definecolor{dfp1}{RGB}{27,158,119}
\definecolor{dfp2}{RGB}{166,206,227}
\definecolor{dfp6}{RGB}{217,95,2}
\definecolor{dfp4}{RGB}{230,171,2}
\definecolor{dfp5}{RGB}{117,112,179}
\definecolor{dfp3}{RGB}{231,41,138}
\definecolor{dfp7}{RGB}{102,166,30}
\definecolor{dfp8}{RGB}{166,118,29}
\definecolor{dfp9}{RGB}{102,102,102}
\definecolor{scp1}{RGB}{166,206,227}
\definecolor{scp3}{RGB}{178,223,138}
\definecolor{scp4}{RGB}{51,160,44}
\definecolor{scp5}{RGB}{251,154,153}
\definecolor{scp6}{RGB}{227,26,28}
\definecolor{scp11}{RGB}{255,255,153}
\definecolor{dkp1}{RGB}{173,216,230}
\definecolor{dkp2}{RGB}{139,0,0}

\definecolor{gtex1}{RGB}{166,206,227}
\definecolor{gtex2}{RGB}{31,120,180}

\definecolor{gtex3}{RGB}{105,105,105}
\definecolor{gtex4}{RGB}{51,160,44}
\definecolor{gtex5}{RGB}{251,154,153}
\definecolor{gtex6}{RGB}{227,26,28}
\definecolor{gtex7}{RGB}{106,61,154}
\definecolor{gtex8}{RGB}{255,255,0}
\definecolor{gtex9}{RGB}{202,178,214}
\definecolor{gtex10}{RGB}{106,61,154}

\definecolor{indic1}{RGB}{255,   255,    0}
\definecolor{indic2}{RGB}{31,   120,  180}
\definecolor{indic3}{RGB}{177,    89,  40}
\definecolor{indic4}{RGB}{51,   160,   44}
\definecolor{indic5}{RGB}{160,  32,  240}
\definecolor{indic6}{RGB}{227,   26,   28}
\definecolor{indic7}{RGB}{253,   191,   111}
\definecolor{indic12}{RGB}{177,   89,   40}

\definecolor{zip1}{RGB}{27,  158,  119}
\definecolor{zip2}{RGB}{255,   127,   0}
\definecolor{zip3}{RGB}{153,    50,  204}
\definecolor{zip4}{RGB}{255,   255,    0}
\definecolor{zip5}{RGB}{124,   252,    0}
\definecolor{zip6}{RGB}{24,   116, 205}
\definecolor{zip7}{RGB}{166,   118,   29}
\definecolor{zip8}{RGB}{46,    46,   46}
\definecolor{zip9}{RGB}{255,    48,   48}
\definecolor{zip10}{RGB}{255,   192,  203}

\definecolor{winep1}{RGB}{27,   158,  119}
\definecolor{winep2}{RGB}{217, 95,    2}
\definecolor{winep3}{RGB}{117,   112,  179}

\definecolor{oilp1}{RGB}{27,   158,  119}
\definecolor{oilp2}{RGB}{217, 95,    2}
\definecolor{oilp3}{RGB}{117,   112,  179}
\definecolor{oilp4}{RGB}{231,    41,  138}
\definecolor{oilp5}{RGB}{102,   166,   30}
\definecolor{oilp6}{RGB}{230,   171,  2}
\definecolor{oilp7}{RGB}{166,   118,  29}
\definecolor{oilp8}{RGB}{102,   102,  102}
\definecolor{oilp9}{RGB}{0,    255,   255}
\definecolor{cmp1}{RGB}{31,120,180}
\definecolor{cmp2}{RGB}{51,160,44}
\definecolor{cmp3}{RGB}{227,26,28}
\definecolor{cmp4}{RGB}{106,61,154}
\definecolor{cmp5}{RGB}{0,0,0}
\definecolor{cmp6}{RGB}{190,190,190}
\definecolor{covid1}{RGB}{27,158,119}
\definecolor{covid2}{RGB}{217,95,2}
\definecolor{covid3}{RGB}{117,112,179}
\definecolor{covid4}{RGB}{231,41,138}
\definecolor{covid5}{RGB}{102,166,30}
\definecolor{covid6}{RGB}{230,171,2}
\definecolor{covid7}{RGB}{166,118,29}
\definecolor{covid8}{RGB}{102,102,102}
\definecolor{covid9}{RGB}{165,15,21}
\definecolor{covid10}{RGB}{8,81,156}

\usepackage{amsthm}

\newcommand{\citep}{\cite}
\newcommand{\citet}{\cite}
\newcommand{\citeyear}{\cite}
\newcommand{\citeauthor}{\cite}

\newtheorem{thm}{Theorem}

\newtheorem{rem}[thm]{Remark}
\newtheorem{res}[thm]{Result}

\usepackage{algorithm,algorithmic}

\usepackage{array,booktabs,longtable,tabularx}
\newcolumntype{L}{>{\raggedright\arraybackslash}X}
\usepackage{siunitx}
\usepackage{caption}

\usepackage{bm}
\usepackage{amsfonts}

\definecolor{ggplot1}{HTML}{E495A5}
\definecolor{ggplot2}{HTML}{BDAB66}
\definecolor{ggplot3}{HTML}{65BC8C}
\definecolor{ggplot4}{HTML}{55B8D0}
\definecolor{ggplot5}{HTML}{C29DDE}

\newcommand{\norm}[1]{\Vert#1\Vert}

\newcommand{\bPsi}{\boldsymbol{\Psi}}

\newcommand{\bOmega}{\boldsymbol{\Omega}}

\newcommand{\be}{\boldsymbol{e}}

\newcommand{\bu}{\boldsymbol{u}}

\newcommand{\bU}{\boldsymbol{U}}
\newcommand{\bx}{\boldsymbol{x}}
\newcommand{\bX}{\boldsymbol{X}}

\newcommand{\bY}{\boldsymbol{Y}}

\newcommand{\bone}{\boldsymbol{1}}

\newcommand{\mR}{\mathcal R}

\newcommand{\R}{\mathbb{R}}
\newcommand{\mS}{\mathbb{S}}
\newcommand{\B}{\mathbb{B}}
\newcommand{\mP}{\mathbb{P}}

\newcommand{\ben}{\begin{enumerate}}
\newcommand{\een}{\end{enumerate}}

\makeatletter
\newcommand\code{\bgroup\@makeother\_\@makeother\~\@makeother\$\@codex}
\def\@codex#1{{\normalfont\ttfamily\hyphenchar\font=-1 #1}\egroup}
\makeatother

\begin{document}
%
\title{Fully Three-dimensional Radial Visualization}

\author{{Yifan~Zhu, Fan~Dai
     and Ranjan~Maitra}
\thanks{Y. Zhu and R. Maitra are with the Department of Statistics
at Iowa State University, Ames, Iowa 50011, USA. e-mail:
\{yifanzhu,maitra\}@iastate.edu.}
\thanks{F. Dai is with the Department of Mathematical Sciences at the Michigan Technological University, Houghton,
  Michigan 49931, USA. e-mail: fand@mtu.edu.}
\thanks{  A portion of this manuscript won Y. Zhu a 2021 Student 
Paper Competition award from the American Statistical Association
(ASA) Section on Statistical Computing and Graphics.
}}

 \IEEEcompsoctitleabstractindextext{
   \begin{abstract}

  We develop methodology for three-dimensional (3D) radial
  visualization (RadViz) of multidimensional datasets. The classical
  two-dimensional (2D) RadViz visualizes multivariate data in the 2D
  plane by mapping every observation to a point inside the unit
  circle. Our tool, RadViz3D, distributes anchor points uniformly on
  the 3D unit sphere. We show that this uniform distribution provides
  the best visualization with minimal artificial visual correlation
  for data with uncorrelated variables. However, anchor points can be
  placed exactly equi-distant from each other only for the five
  Platonic solids, so we provide equi-distant anchor points for these
  five settings, and approximately equi-distant anchor points via a
  Fibonacci grid for the other cases. Our methodology, implemented in
  the R package {\tt radviz3d}, makes fully 3D RadViz possible and is
  shown to improve the ability of  this nonlinear technique in more
  faithfully displaying simulated data as well as the crabs, olive
  oils and wine datasets. Additionally, because radial visualization
  is naturally suited for  compositional data, we use  RadViz3D to
  illustrate (i) the chemical composition of Longquan celadon ceramics and their Jingdezhen imitation over centuries, and (ii) US regional SARS-Cov-2 variants' prevalence in the Covid-19 pandemic during the summer 2021 surge of the Delta variant.

\end{abstract}

%
 }
\maketitle
 \IEEEdisplaynotcompsoctitleabstractindextext
%
\vspace{-0.2em}
\section{Introduction}
\label{sec:intro}
%
%
%
%
Graphical display of multivariate data is important to obtain insight into
their properties and similarity or distinctiveness of different
groups~\citep{cardetal99}. The goal of effective visualization is to
map multi-dimensional observations  to a
lower-dimensional space, with the reduced display conveying
information on the characteristics as faithfully  as possible.

Continuous multivariate data are    displayed in many ways~\citep{bertinietal11,fonnetandprie21}
({\em e.g.} starplots
\allowbreak
~\citep{chambersetal83}, Chernoff 
faces~\citep{chernoff73}, parallel coordinate
plots~\citep{inselberg85,wegman90}, surveyplots~\citep{fayyadetal01}, Andrews'
curves~\citep{andrews72,khattreeandnaik02}, biplots~\citep{gabriel71}, star coordinate
plots~\citep{kandogan01},
the grand tour~\citep{asimov85,bujaetal05}, Uniform Manifold Approximation and Projections
(UMAP)~\citep{mcinnesetal18}), but our focus in this short technical
note is on improving the nonlinear display called
radial visualization or
RadViz~\citep{hoffmanetal97,hoffmanetal99,grinsteinetal01,draperetal09}
that projects data onto a circle using Hooke's law. Here, 
$p$-dimensional observations are projected onto the 2D plane using $p$ anchor
points arranged to be on the perimeter of a circle. This
representation places each observation at the center of the circle
that is then pulled by springs in the directions of the $p$ anchor
points while being balanced by forces relative to the coordinate
values. Observations with similar relative values across all
attributes are placed close to the center while the others are
closer to anchor points corresponding to the coordinates with
higher relative values.  However, there is loss of
information~\citep{arteroanddeoliveira04} in RadViz which maps a
$p$-dimensional point to 2D.
This loss worsens with increasing $p$, but may potentially be
alleviated by extending it to 3D.
The Viz3D approach \citep{arteroanddeoliveira04} extends 2D RadViz (henceforth, RadViz2D) by simply adding to the 2D projection a third
dimension that, for each observation, is simply the average of all its
attributes. The extension robs Viz3D of its interpretability in terms of
the attributes. Its improvement over RadViz2D is also
by design limited, so here we investigate the possibility of
developing a truly 3D extension of RadViz. 

One challenge of extending RadViz from two  to three dimensions is
that a 3D sphere can be exactly divided into $p$  regions of
equal volumes only for the Platonic solids. However,
Section~\ref{sec:method} shows that placing equi-spaced anchor points is
necessary to reduce artificially induced 
correlation between variables. So, for other $p$,
Section~\ref{sec:method} uses a Fibonacci grid method to develop an
approximate solution. We call our method RadViz3D
and illustrate in Section~\ref{sec:illustration} its ability to more
accurately display structure in simulated and some common real
datasets, than RadViz2D or Viz3D. Further, because radial
visualization is, by construction, naturally suited for compositional
data, we use RadViz3D on two such datasets. In the first case, we
illustrate the composition, over history, of celadon ceramics from
Longquan kilns and that of  their Jingdezhen kiln-manufactured imitations. Our second example uses RadViz3D to
illustrate the changing face of the Covid-19 pandemic in the ten
Health and Human Services (HHS) regions of the United 
States (US) over the course of the summer of 2021, in terms of the
proportion of the variants  ofthe Severe Acute Respiratory Syndrome
Coronavirus~2 (SARS-Cov-2).
The main paper concludes with some discussion. We also have an online supplemental
HTML resource (also at 
\url{https://radviz3d.github.io})
that allows the reader to more fully experience the benefits of
RadViz3D (and Viz3D) on our
examples. 

\section{Methodology}
\label{sec:method}
\subsection{Generalized radial visualization}
\label{subsec:gradviz}
We naturally extend  the classic RadViz2D by defining generalized
radial visualization (GRadViz) as a map of $\bX =
(X_1,X_2,\ldots,X_p)^\top\in \R^p$ to a point in $\mS^{q} = \{\bx \in \mR^{q+1}: 
\|\bx\| = 1\}$ using
\begin{equation}
  \bPsi(\bX; \bU)= \frac{\bU \bX}{\bone_p^\top\bX},
\label{eq:T}
\end{equation}
where $\bone_p = (1,1,\ldots,1)^\top$, 
and $\bU = [\bu_1\vdots\bu_2\vdots\ldots\vdots\bu_p]$ is a projection
matrix with $j$th column and anchor point  $\bu_j\in\mS^q$,
 for $j
= 1, 2, \ldots, p$. For $q=1$, GRadViz reduces to RadViz2D.

\begin{rem}
  As helpfully suggested by a reviewer, the use of $\bX/\bone^\top\bX$  in
  \eqref{eq:T} means that it is ideally set up for compositional
  data. We illustrate performance with such data in~Sections~\ref{sec:ceramics} and \ref{sec:application}.
\end{rem}

As in RadViz2D, our generalization $\bPsi(\cdot;\cdot)$ also has a physical
interpretation. Suppose that we have $p$ springs connected to the 
anchor points $\bu_1, \bu_2, \ldots, \bu_p \in \mS^q$ and that these $p$ springs have  
spring constants $X_1, X_2, \ldots, X_p$. Let $\bY\in\mR^{q+1}$ be
the equilibrium point of the system. Then we have
 $ \sum_{j=1}^p X_j (\bY - \bu_j) = 0,$
with our generalization $\bY= \bPsi(\bX; \bU)$ as its solution. 

GRadViz is actually a special case of  normalized radial
visualization (NRV) \citep{danielsetal12} that allows the anchor
points to lie outside the hypersphere and is point-ordering-,
line-ordering-, and 
convexity-invariant. These desirable properties for visualization
are also inherited by $\bPsi(\cdot;\cdot)$. 
However, GRadViz is 
scale-invariant, that is, $\bPsi(k \bX; \bU)\! =\! \bPsi(\bX; \bU)$ for
any $k\neq 0$. In other words, any line passing through the origin is
projected to a single point in the radial visualization. So the
visualization is invariant under scaling, and only the relative values
of the coordinates matter. This means that the coordinates of the
dataset have to be comparable, so we use 
the minmax transformation on the $j$th feature ($j=1,2,\ldots,p$) of
the $i$th observation that is given by
\begin{equation}
  m_j(X_{ij}) = \frac{X_{ij} - \min_{1\leq i \leq n} X_{ij}}{\max_{1\leq
      i \leq n} X_{ij} - \min_{1 \leq i \leq n} X_{ij}}.
\label{eq:minmax}
\end{equation}
The minmax transformation places every observation in $[0,1]^{p}$,
ensuring that the data, after also applying  
$\bPsi(\cdot;\cdot)$, are all inside the unit ball $\B^{q+1} =
\{\bx\in\mR^{q+1}:\norm{x}\leq 1\}$. Outliers, if any, are
placed close to the boundary of the ball. Other standardization
teciniques like centering and scaling do not enjoy these properties,
so we use the minmax transformation as the default standardization for
this article.

The placement of the anchor points is another issue to be addressed in GRadViz, with
different points yielding very different visualizations. Now
suppose that $\bX$ has  $p$ uncorrelated coordinates. For any $\bX_i, \bX_j \in {\mR}^p$, let $\bY_l = \bPsi(\bX_l; \bU), l\in\{i,j\}$ be 
the GRadViz-transformed data. The squared Euclidean distance between
$\bY_i$ and $\bY_j$ is
\begin{equation*}
\norm{\bY_i - \bY_j}^2 = \left(\frac{\bX_i}{\bone_p^\top \bX_i} -
  \frac{\bX_j}{\bone_p^\top\bX_j}\right)^\top\bU^\top \bU \left(\frac{\bX_i}{\bone_p^\top \bX_i} - \frac{\bX_j}{\bone_p^\top \bX_j}\right),
\end{equation*}
a quadratic form with positive definite matrix
$\bU^\top\bU$ that has the 
$l$th diagonal element $\bu_l^\top\bu_l=1$ and  $(k,l)$th entry
$\bu_{k}^\top\bu_l = \cos\langle\bu_k, \bu_l\rangle$.
For $\bX_l = a_l\be_l$, with $\be_l$ 
as the $l$th standard unit vector ($l \in\{ i,j\}$),
\begin{equation}
  \norm{\bY_i - \bY_j}^2 = 2 - 2\cos(\bu_i, \bu_j).
  \label{eq:dist}
\end{equation}
The $i$th and $j$th coordinates of $\bX_i$ and $\bX_j$ in this example
are as discordant 
  as possible from each other (which is very likely to happen if $i$th and $j$th coordinates are negatively
  correlated or uncorrelated), playing an important role in identifying distinctive data points, and should be placed as far apart as
  possible (in opposite directions) in the radial visualization.
  However~\eqref{eq:dist} shows that the squared Euclidean distance between $\bY_i$ and $\bY_j$ approaches 0 as the angle between
  $\bu_i$ and $\bu_j$ approaches 0. Therefore, the radial visualization will display such points to be more similar and fails to acknowledge the differences in the $i$th and the $j$th coordinates if the angle between  $\bu_i$    and
  $\bu_j$ is less then $\pi/2$, which can
  create artificial positive visual correlation between the $i$th and $j$th
  coordinates (positive visual correlation here means that visualization will be like the case when the $i$th and the $j$th coordinates are positively correlated).  Since not all pairs of coordinates will be negatively correlated, when two coordinates are positively correlated, they are less important in differentiating distinctive data points and therefore  positively correlated coordinates should be placed  close
  together so that there is more room to place negatively correlated coordinates apart. When all coordinates are uncorrelated, they are equally important in differentiating distinctive data points. To reduce such effects, we need to
  distribute the anchor points as far away from each other as
  possible. Therefore, our GRadViz formulation recommends evenly-distributed anchor points on   $\mS^q$. RadViz3D has an inherent advantage over RadViz2D  because it can more readily facilitate larger angles 
  between anchor points as the smallest angle between any two of $p$ (fixed)
  evenly-distributed anchor points in RadViz3D 
  is always larger than that in 
  RadViz2D.
 (Indeed, higher-dimensional displays beyond
  3D  would be even more beneficial were it possible to obtain such
  displays.)  
  For example, with $p = 4$, we can place anchor points in 3D so
  that  the   angle between any two of them is the same, but this is
  not possible in 2D.  
  At the same time, RadViz2D can not place
  multiple positively correlated coordinates next to each other, as desirable for accurate 
  visualization~\citep{ankerstetal96}. The placement of anchor points
  therefore has more pronounced importance in
  RadViz2D~\citep{dicaroetal10,sharkoetal08} than  RadViz3D.

\begin{rem}
  Our discussion on GRadViz provides the rationale behind
  radial visualization with equi-spaced anchor points. It also shows that
  investigating spacings between anchor points, as 
  done~\citep{dicaroetal10,vanlongandngan15,chengetal17} for RadViz2D,
  is unnecessary and perhaps even misleading because of its
  potential to induce artificial visual correlation between
  coordinates. However, as pointed out by a reviewer, investigating
  layouts and arrangements of the axes is still  important for cases with $p>3$
  in RadViz2D, and $p>4$ for RadViz3D.
 
\end{rem}

\subsection{Three-dimensional radial visualization}
\label{subsec:radviz3d}
Following the setup and discussion in
Section~\ref{subsec:gradviz}, let $\bPsi: \mathbb{R}^{p} \mapsto
\B^{3}=\{\bx\in{{\R}^3}:\norm{\bx}\leq 1\}$ map $\bX\in\R^p$ to $\bPsi(\bX;\bU) = \bU\bX/\bone_p^\top\bX$ 
with $\bU$ as before and with $j$th column (anchor point) $\bu_j$
that, we have contended, should be as evenly-spaced in $\mS^2$ 
as possible.  We find the set $\wp$ of
equi-spaced anchor points $\bu_1,\bu_2,\ldots,\bu_p$ from Result~\ref{res}.
\begin{res}
  \label{res}
{\em Anchor Points Set}. Denote the golden ratio by $\varphi =
  (1+\sqrt5)/2$. For $p\in\{4,6,8,12,20\}$, the elements in $\wp$
have the coordinates listed in Table~\ref{tab:anchor}.
%
%
\begin{table}[h]
\caption{Anchor points set for  $p\!\in\!\{4,6,8,12,20\}$. Here $\varphi\!=\!(1\!+\!\sqrt5)/2$.} 
\label{tab:anchor}
\resizebox{\textwidth}{!}{ 
\resizebox{.5\textwidth}{!}{\begin{tabular}{ |r||c|c| } \hline
$p$ &Platonic Solid & $\wp$ \\ \hline
4 & Tetrahedron & $\{(1,1,1)/\sqrt3, (1,-1,-1)/\sqrt3,(-1,1,-1)/\sqrt3,(-1,-1,1)/\sqrt3\}$\\ \hline
6 & Octahedron & $\{(\pm 1, 0, 0), (0,\pm 1, 0), (0, 0, \pm 1)\}$ \\
  \hline
8 & Cube & $\{\pm 1, \pm 1, \pm 1\}$ \\ \hline
12 & Dodecahedron & $\{ (0,\pm1,\pm\varphi), (\pm1,\pm\varphi,0) (\pm\varphi,0,\pm1)\}/\sqrt{1+\varphi^2}$\\ \hline
20 & Icosahedron & $\{ (\pm1,\pm 1,\pm1)/\sqrt3, (0,\pm\varphi^{-1},\pm\varphi)/\sqrt3, (\pm\varphi^{-1},\pm\varphi,1)/\sqrt3,
    (\pm\psi,0,\pm\varphi^{-1})/\sqrt3\}$\\ \hline
                             \end{tabular}}
                         }
    \end{table}
For other integers $p\geq5$, only an approximate solution is possible:
here the elements of $\wp$ are $\bu_j^\top=(u_{j1},u_{j2},u_{j3}), j =1,2,\ldots,p$ with
$u_{j1} = \cos(2\pi j \varphi^{-1})(1-u_{j3}^2)^{-\frac12},
u_{j2} = \sin(2\pi j \varphi^{-1})(1-u_{j3}^2)^{-\frac12},
u_{j3} = (2j-1)/p-1.$
 \begin{proof}
   For $p\in\{4,6,8,12,20\}$, the coordinates are exactly equi-spaced with anchor points
   corresponding to the vertices of the Platonic solids. For other values of $p\geq 5$,  
 we derive an approximate solution by implementing a Fibonacci grid
 method \citep{gonzalez10} that produces the latitude $\phi_j$ and
 longitude $\theta_j$ of the $j$th anchor point on $\mS^3$ as
 $\phi_j = \arcsin a_j, \theta_j = 2\pi j \varphi^{-1}$
 with $a_1,a_2,\ldots,a_p$ an arithmetic progression chosen 
 to have common difference $\displaystyle{{2}/{p}}$. We take $a_1 =
 \displaystyle{{1}/{p}-1}$, so $a_j = {2}(j-1)/p+a_1 = ({2j-1})/{p}-1$.
 The  Cartesian coordinates of the $j$th anchor point
 $\bu_j\in\wp$ follow by transforming between coordinate systems. 
 \end{proof}
%
%
\end{res}
 \begin{rem}
 The geometric solutions of $\wp$ for $p\in\{4,6,8,12,20\}$ are closely
 related to the Thomson problem in traditional molecular quantum
 chemistry~\citep{atiyahandsutcliffe03}.
  Also, for $p\geq5$, $p\not\in\{4,6,8,12,20\}$, 
   the approximate solution 
   distributes anchor points along a generative spiral on $\mS^2$, 
   with consecutive points as separated from each other as possible, satisfying the ``well-separation''
   property \citep{saffandkuijlaars97}. 
 \end{rem}
Result~\ref{res} provides the wherewithal for RadViz3D for $p\geq4$ by
projecting each observation $\bX_i\in \R^p,i=1,2,\ldots,n$ 
to $\bPsi(\bX_i;\bU = [\bu_1\vdots\bu_2\vdots\ldots\vdots \bu_p])$.
RadViz3D displays of multidimensional data can then
be obtained via 3D graphics to facilitate the discovery of patterns and structure. 
\section{Illustrative Examples}
 \label{sec:illustration}
 We illustrate RadViz3D on some moderately-dimensioned simulated and
standard classification  datasets and show that it can reveal distinctiveness of
class and other structure better than RadViz2D or Viz3D. Because
RadViz3D and Viz3D are 3D displays, we only show some of their static
views in this paper, referring the reader to the HTML resource for
their fuller appreciation. Further, we use the overlap heatmap
of $\bOmega$ to display the numerical 
distinctiveness of each class relative to the others. We define
$\bOmega$ in the following way. Let $\omega_{j\mid i}$ be the
probability of misclassifying an observation from the $i$th group into the
$j$th group. Then $\bOmega$ is the symmetrized matrix with $(i,j)$th
element consisting of the pairwise overlap~\citep{maitraandmelnykov10}
$\omega_{ij}=\omega_{j\mid i} +\omega_{i\mid j}$. The overlap heatmap
displays $\omega_{ij}$ for $i>j$.
 \subsection{Simulated datasets}
We simulated 5D datasets of $n=500$ observations from five groups of known group
separation and clustering complexity. The \texttt{MixSim} 
package~\citep{melnykovetal12} in R~\citep{R} allows for the simulation of class data according to a pre-specified {\em
  generalized overlap} ($\ddot\omega$) measure~\citep{melnykovandmaitra11} that
indexes clustering complexity, with a
\begin{figure*}[!h]
  \vspace{-0.35in}
   \mbox{
     \setcounter{subfigure}{-4}
     \subfloat[$\ddot\omega=0.001$]{
       \begin{minipage}[b][][t]{\textwidth} 
      \mbox{
      \subfloat{\includegraphics[width=.25\textwidth]{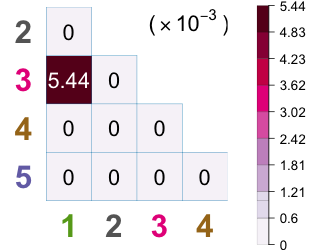}}
      \subfloat{\includegraphics[width=.25\textwidth]{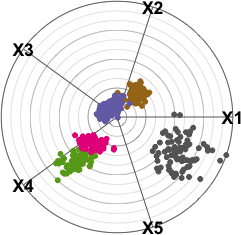}}
        \subfloat{\includegraphics[width=.24\textwidth]{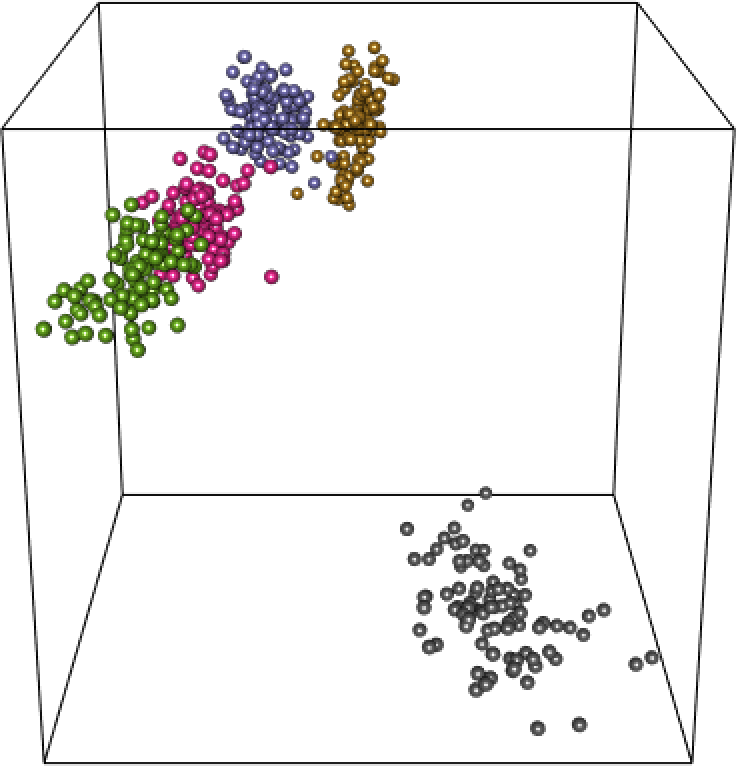}}
        \subfloat{\includegraphics[width=.25\textwidth]{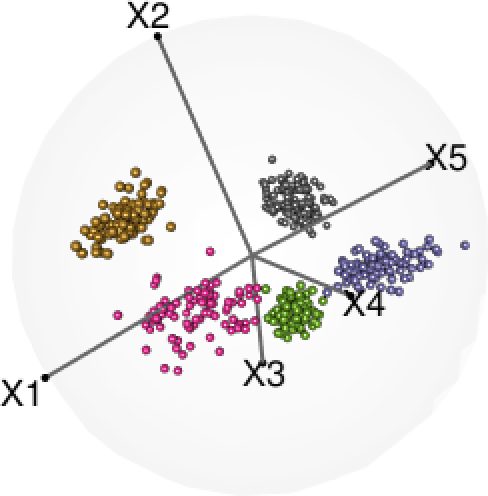}}
      }
\vspace{-0.2in}
      \end{minipage}}%
  }
   \mbox{
     \setcounter{subfigure}{-3}
     \subfloat[$\ddot\omega=0.01$]{
       \begin{minipage}[b][][t]{\textwidth} 
      \mbox{
      \subfloat{\includegraphics[width=.25\textwidth]{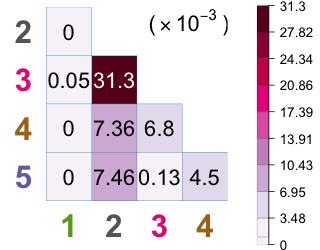}}
      \subfloat{\includegraphics[width=.25\textwidth]{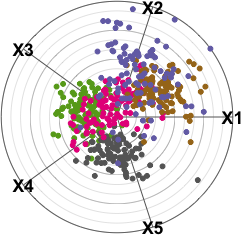}}
        \subfloat{\includegraphics[width=.24\textwidth]{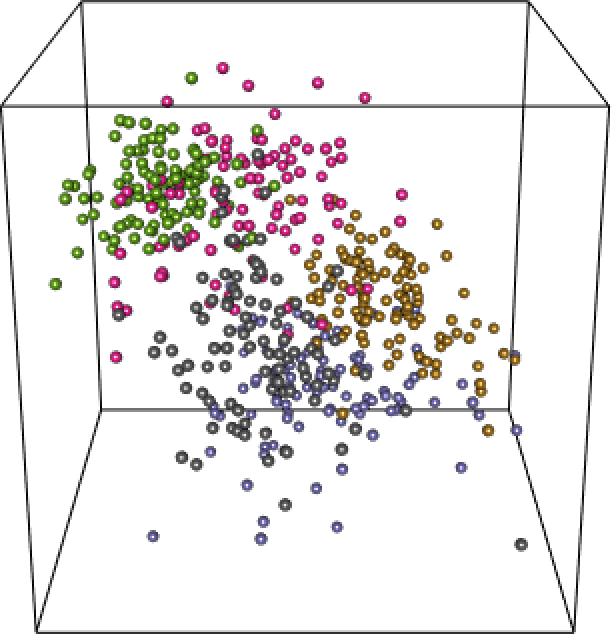}}
        \subfloat{\includegraphics[width=.25\textwidth]{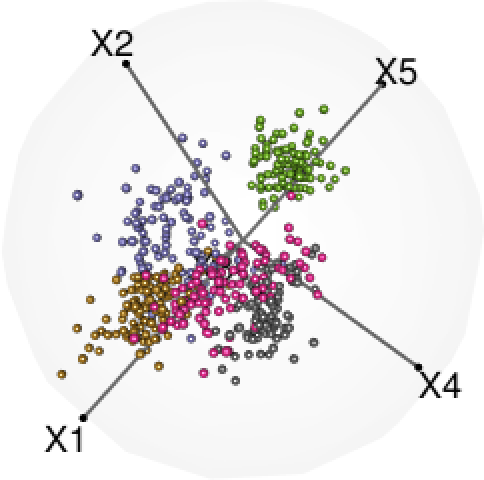}}
      }
\vspace{-0.2in}
      \end{minipage}}%
\vspace{-0.2in}
  }
   \mbox{
     \setcounter{subfigure}{-2}
     \subfloat[$\ddot\omega=0.05$]{
       \begin{minipage}[b][][t]{\textwidth} 
      \mbox{
      \subfloat{\includegraphics[width=.25\textwidth]{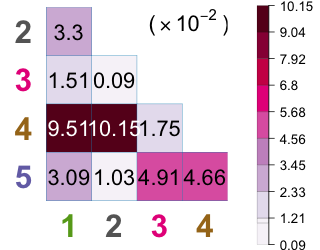}}
      \subfloat{\includegraphics[width=.25\textwidth]{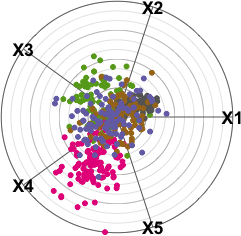}}
        \subfloat{\includegraphics[width=.24\textwidth]{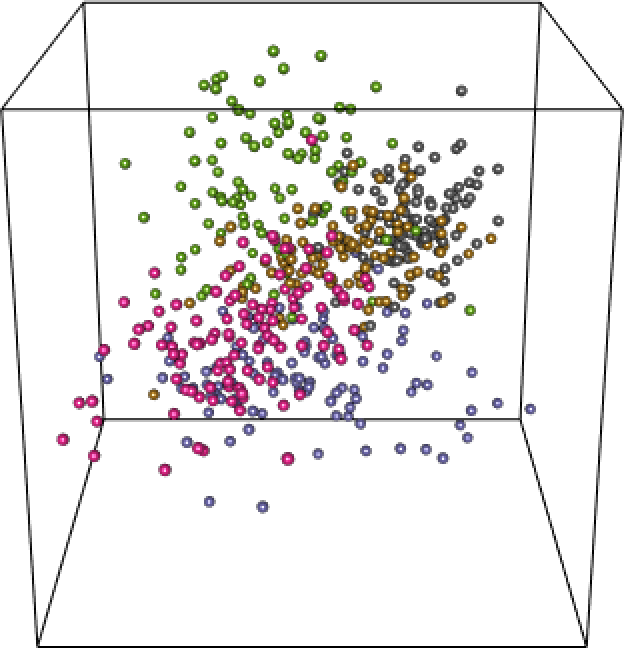}}
        \subfloat{\includegraphics[width=.25\textwidth]{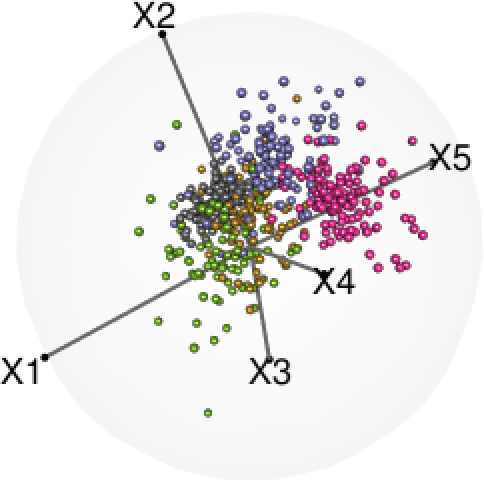}}
      }
      \end{minipage}}%
  }
\vspace{-.25cm}
\caption{Overlap heatmaps of $\bOmega$ and RadViz2D, Viz3D and RadViz3D displays of simulated 5D datasets of varying clustering complexity ($\ddot
  \omega$).  Colors of class labels are the same for each setting.}
\label{fig:simcdr}
\vspace{-.25cm}
\end{figure*}
small value ($\ddot\omega\!\!=\!\!10^{-3}$) implying  good separation between groups and larger values
($\ddot\omega = 0.01, 0.05$) indicating increasingly poorer separation
and increased overlap. Fig.~\ref{fig:simcdr} displays the results and shows that the \texttt{MixSim}-generated pairwise overlaps between the 5 groups are essentially preserved in the RadViz3D display. For
$\ddot\omega\!\!=\!\!10^{-3}$ (Fig.~\ref{fig:simcdr}a), the relatively
higher overlap between Groups~1 and 3 is captured across all three
displays, but unlike RadViz3D, both RadViz2D and Viz3D place both Groups~4 and 5 fairly
close together, belying their good separation ($\omega_{45}\approx
0$). The more appropriate use of the third dimension by RadViz3D than
by Viz3D provides greater benefit at higher $\ddot\omega$. For
instance, Viz3D and RadViz2D show little qualitative  difference in
the distinctiveness of the groups between Figs.~\ref{fig:simcdr}b and
\ref{fig:simcdr}c, but RadViz3D 
shows that while the five classes are less separated in
Fig.~\ref{fig:simcdr}b than in Fig.~\ref{fig:simcdr}a, the groups are
still fairly distinguishable here relative to Fig.~\ref{fig:simcdr}c. 
Therefore, the RadViz3D display provides a more accurate
and meaningful representation of the known class structure of these datasets. 
\subsection{Some common real-data examples}
\label{sec:app}
We now visualize three multivariate datasets often 
used in the statistical graphics literature. 
\subsubsection{Crabs} 
\begin{figure}
\centering
\mbox{
\subfloat[RadViz2D]{\label{crabs:2d}\includegraphics[width=0.25\textwidth]{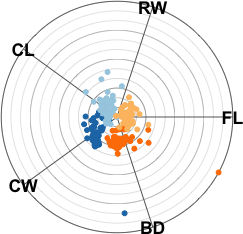}}
\subfloat[Viz3D]{\label{crabs:viz3d}\includegraphics[width=0.24\textwidth]{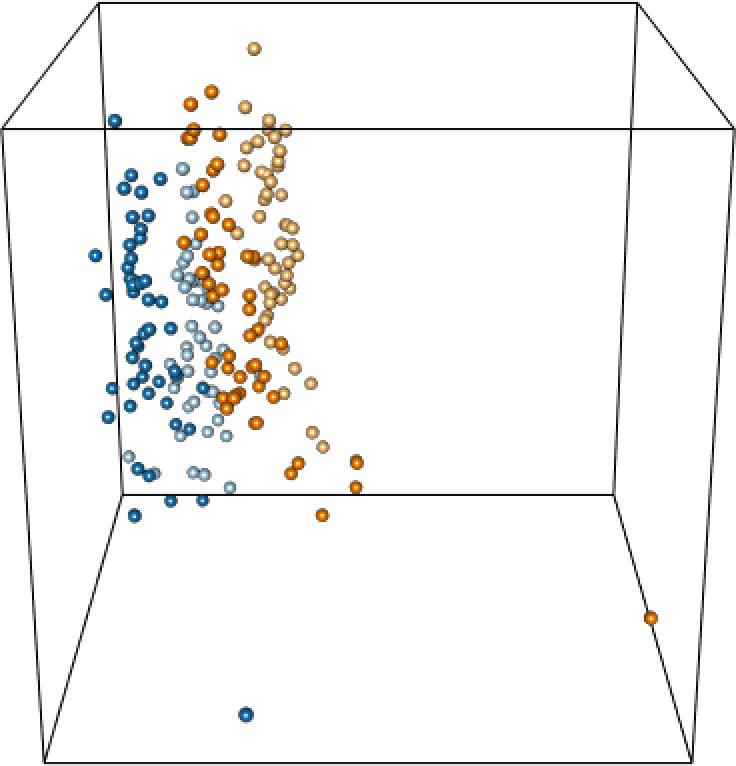}}
\subfloat[RadViz3D]{\label{crabs:radviz3d}\includegraphics[width=0.25\textwidth]{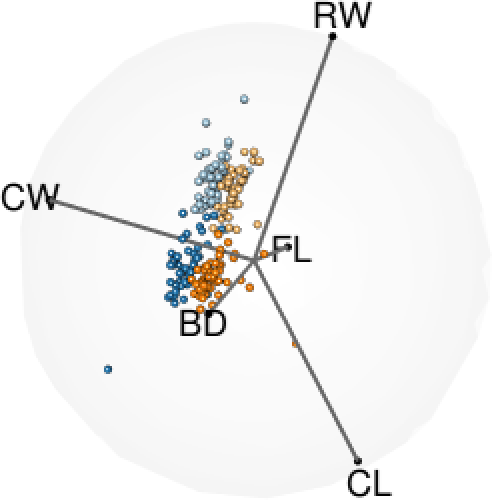}}
\subfloat[Overlap heatmap]{\label{crabs:overlap}\includegraphics[width=0.25\textwidth]{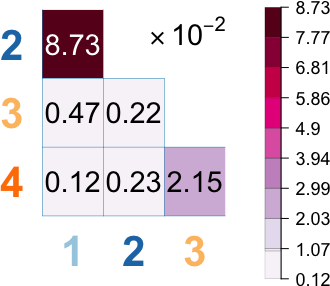}}
}
\centering
\text{
\tikz\draw[cbp1,fill=cbp1] (0,0) circle (.5ex); Male-Blue 
\tikz\draw[cbp2,fill=cbp2] (0,0) circle (.5ex); Female-Blue 
\tikz\draw[cbp3,fill=cbp3] (0,0) circle (.5ex); Male-Orange 
\tikz\draw[cbp4,fill=cbp4] (0,0) circle (.5ex); Female-Orange}
\caption{RadViz2D, Viz3D, RadViz3D displays and the overlap heatmap of the
  {\em Crabs} dataset.}
\label{crabs}
\end{figure}

\begin{figure}
\centering
\mbox{
\subfloat[RadViz2D]{\label{crabs-ncl:2d}\includegraphics[width=0.25\textwidth]{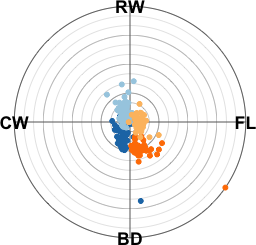}}
\subfloat[Viz3D]{\label{crabs-ncl:viz3d}\includegraphics[width=0.24\textwidth]{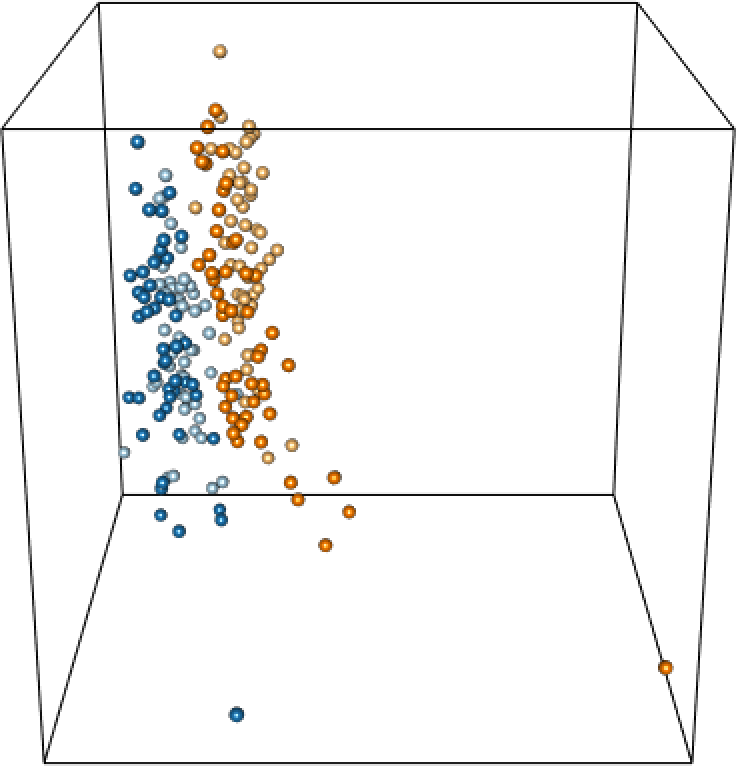}}
\subfloat[RadViz3D]{\label{crabs-ncl:radviz3d}\includegraphics[width=0.25\textwidth]{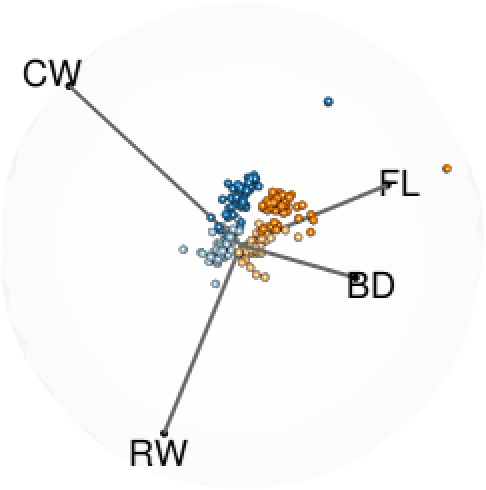}}
\subfloat[Overlap Map]{\label{crabs-ncl:overlap}\includegraphics[width=0.25\textwidth]{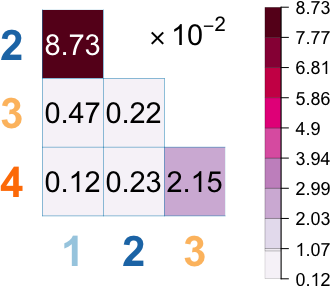}}
}
\centering
\text{
\tikz\draw[cbp1,fill=cbp1] (0,0) circle (.5ex); Male-Blue 
\tikz\draw[cbp2,fill=cbp2] (0,0) circle (.5ex); Female-Blue 
\tikz\draw[cbp3,fill=cbp3] (0,0) circle (.5ex); Male-Orange 
\tikz\draw[cbp4,fill=cbp4] (0,0) circle (.5ex); Female-Orange}
\caption{Radial visualization displays and overlap heatmap of the
  {\em Crabs} dataset sans CL.}
\label{crabs-ncl}
\end{figure}

This dataset~\citep{campbell1974multivariate} has measurements on morphological
characteristics (FL or frontal lip width, RW or rear width, CL or
carapace midline length, CW or carapace maximum width, BD or body
depth) of 50 crabs each with blue or orange shells and of each
gender. Fig.~\ref{crabs} shows the three displays along with the
 overlap heatmap of $\bOmega$. 
We see less distinction between the males and females as
per $\bOmega$, but that is not quite captured in the RadViz2D displays (and
definitely not so in Viz3D). However, the RadViz3d display indicates that CL may be redundant in the display, as also reported
in~\citet{rafteryanddean06}. Removing this variable results in a
clearer display~(Fig.~\ref{crabs-ncl}) where the RadViz3D display shows not just
the higher overlap between the blue males and females relative to
those between the orange males and females, but also that the species
are less distinguishable, in terms of these characteristics, in the
males than in the females. 
In this illustration, RadViz3D is seen to alone provide important
information on the structure of the data in terms of identifying both
the redundancy of CL as well as faithful distinction between the
classes.

\subsubsection{Wine} 
This dataset~\citep{forinaetal88} is on the
chemical composition of 178 wines from three (Barolo, Gringolino and
Barbera) cultivars 
\begin{figure}
\centering
\mbox{
\subfloat[RadViz2D]{\label{wine:radviz2d}\includegraphics[width=0.33\textwidth]{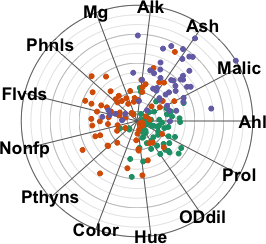}}
\subfloat[Viz3D]{\label{wine:viz3d}\includegraphics[width=0.33\textwidth]{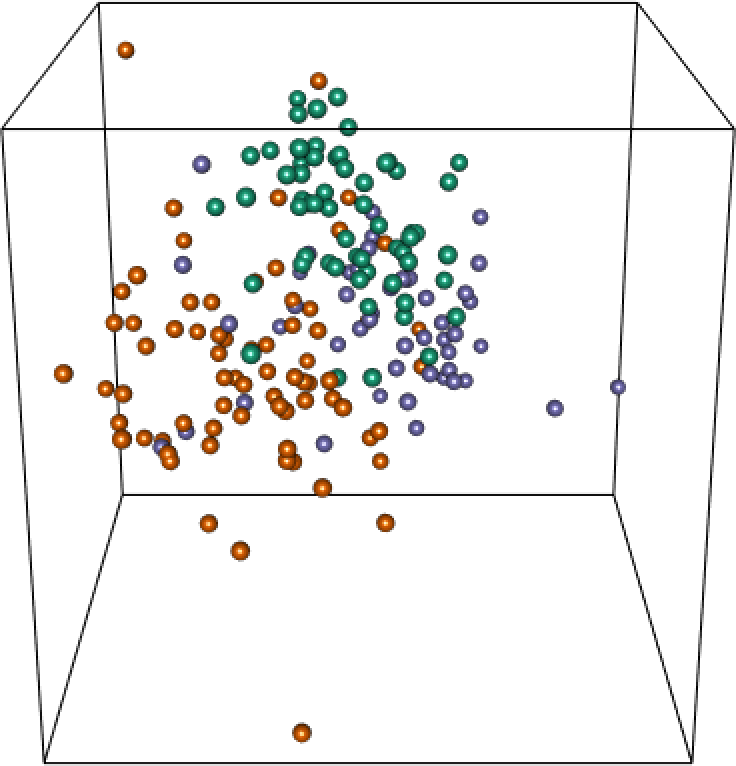}}
\subfloat[RadViz3D]{\label{wine:radviz3d}\includegraphics[width=0.33\textwidth]{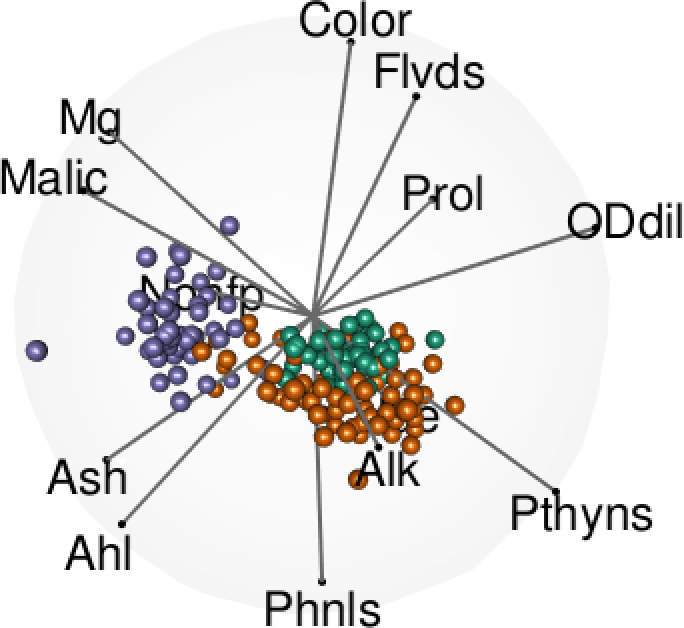}}}
\centering
\text{
\tikz\draw[winep1,fill=winep1] (0,0) circle (.5ex); Cultivar 1 
\tikz\draw[winep2,fill=winep2] (0,0) circle (.5ex); Cultivar 2 
\tikz\draw[winep3,fill=winep3] (0,0) circle (.5ex); Cultivar 3}
\caption{RadViz2D, Viz3D and RadViz3D displays of the {\em Wine}  dataset.}
\label{wine}
\vspace{-0.25in}
\end{figure}
of grapes grown in the same region of Italy. Fig.~\ref{wine} provides
the three displays of the dataset through the composition of thirteen chemicals: Alcohol (Ahl),  Malic acid (Malic),
Ash, Alkalinity of ash (Alk), Magnesium (Mg), Total phenols (Pnls),
Flavanoids (Flvds), Nonflavanoid phenols (Nonfp), Proanthocyanins
(Pthyns), Color intensity, Hue, OD280/OD315 of diluted wines (ODdil)
and Proline (Prol).  All three displays can not correctly distinguish
the first two cultivars from each other but RadViz3D alone separates
the third cultivar from the others very well, more accurately reflecting the
overlap measures $(\omega_{13}\approx0, \omega_{23}\approx 9\times 10^{-4})$. 

\subsubsection{Olive oils} This dataset
\citep{forinaandtiscornia82,forinaetal83} provides the composition of
eight fatty (palmitic, palmitoleic, stearic, oleic, linoleic,
linolenic, arachidic, eicosenoic) acids in 572 olive oil samples from
nine regions in three areas of Italy. 
\begin{figure}
\centering
\mbox{
\subfloat[RadViz2D]{\label{oils:2d}\includegraphics[width=0.33\textwidth]{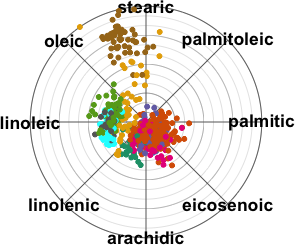}}
\subfloat[Viz3D]{\label{oils:viz3d}\includegraphics[width=0.28\textwidth]{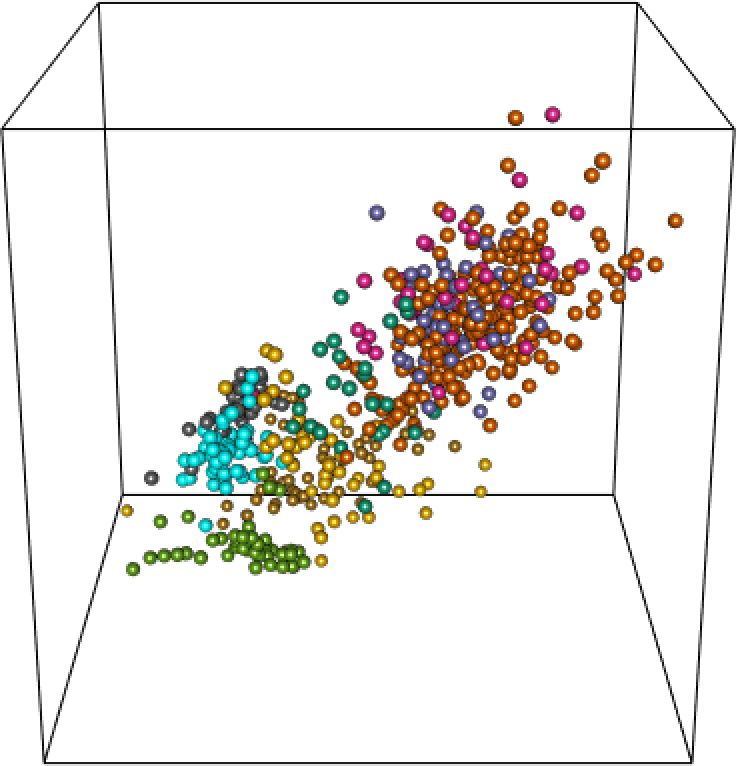}}
\subfloat[Overlap Map]{\label{oils:overlap}\includegraphics[width=0.33\textwidth]{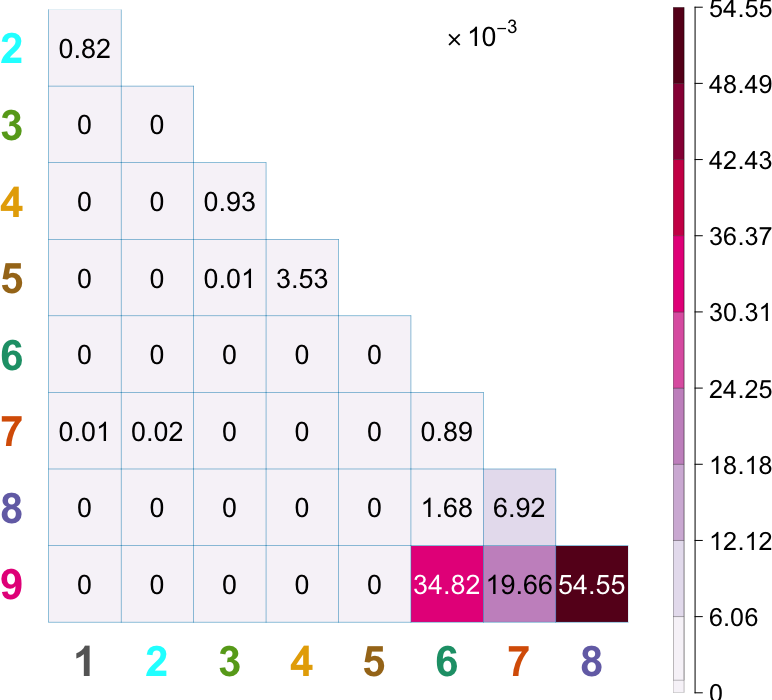}}
}\\
\mbox{
\subfloat[RadViz3D, Sardinia]{\label{oils-sard:radviz3d}\includegraphics[width=0.31\textwidth]{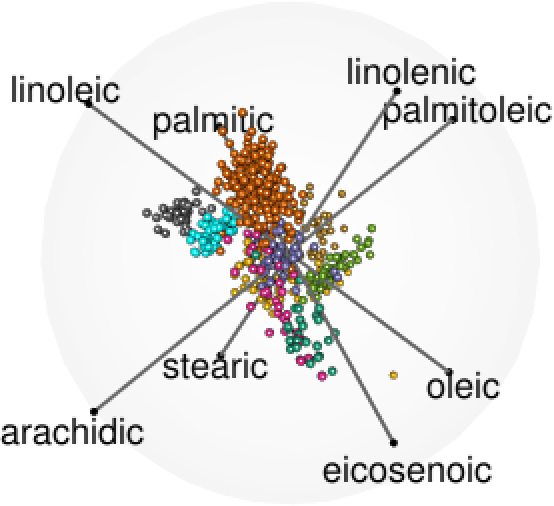}}
  \subfloat[RadViz3D, Northern Italy]{\label{oils-north:radviz3d}\includegraphics[width=0.31\textwidth]{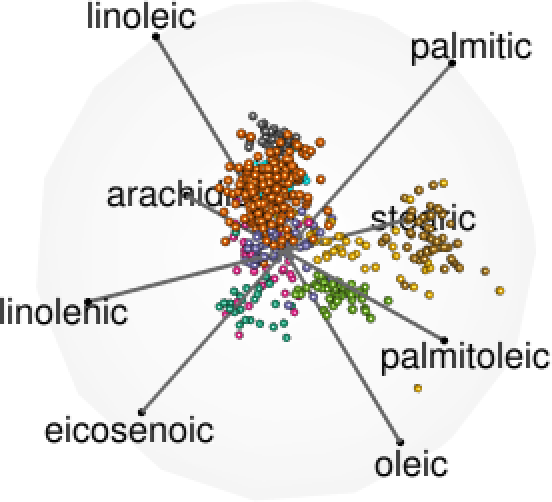}}
\subfloat[RadViz3D, Southern Italy]{\label{oils-south:radviz3d}\includegraphics[width=0.33\textwidth]{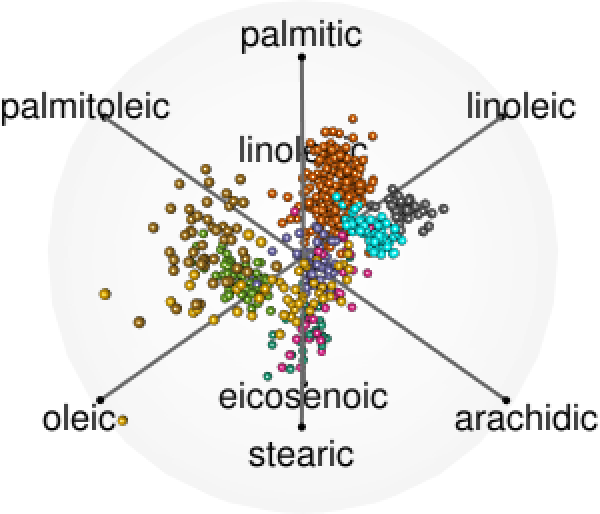}}
}
\centering
\text{
\tikz\draw[oilp8,fill=oilp8] (0,0) circle (.5ex); Coast-Sardinia 
\tikz\draw[oilp9,fill=oilp9] (0,0) circle (.5ex); Inland-Sardinia
\tikz\draw[oilp5,fill=oilp5] (0,0) circle (.5ex);
Umbria
\tikz\draw[oilp6,fill=oilp6] (0,0) circle (.5ex); East-Liguria
\tikz\draw[oilp7,fill=oilp7] (0,0) circle (.5ex); West-Liguria
}\\
\text{
\tikz\draw[oilp1,fill=oilp1] (0,0) circle (.5ex); 
North-Apulia
\tikz\draw[oilp2,fill=oilp2] (0,0) circle (.5ex); South-Apulia
\tikz\draw[oilp3,fill=oilp3] (0,0) circle (.5ex); Calabria
\tikz\draw[dfp4,fill=oilp4] (0,0) circle (.5ex); Sicily
}\\
\caption{Olive oils dataset: (a) RadViz2D and (b) Viz3d displays, and (c) Overlap map
  $\bOmega$. (d-f) RadViz3D displays, each showing the
  distinctiveness of an area's olive oils.}
\label{oils}
\end{figure}
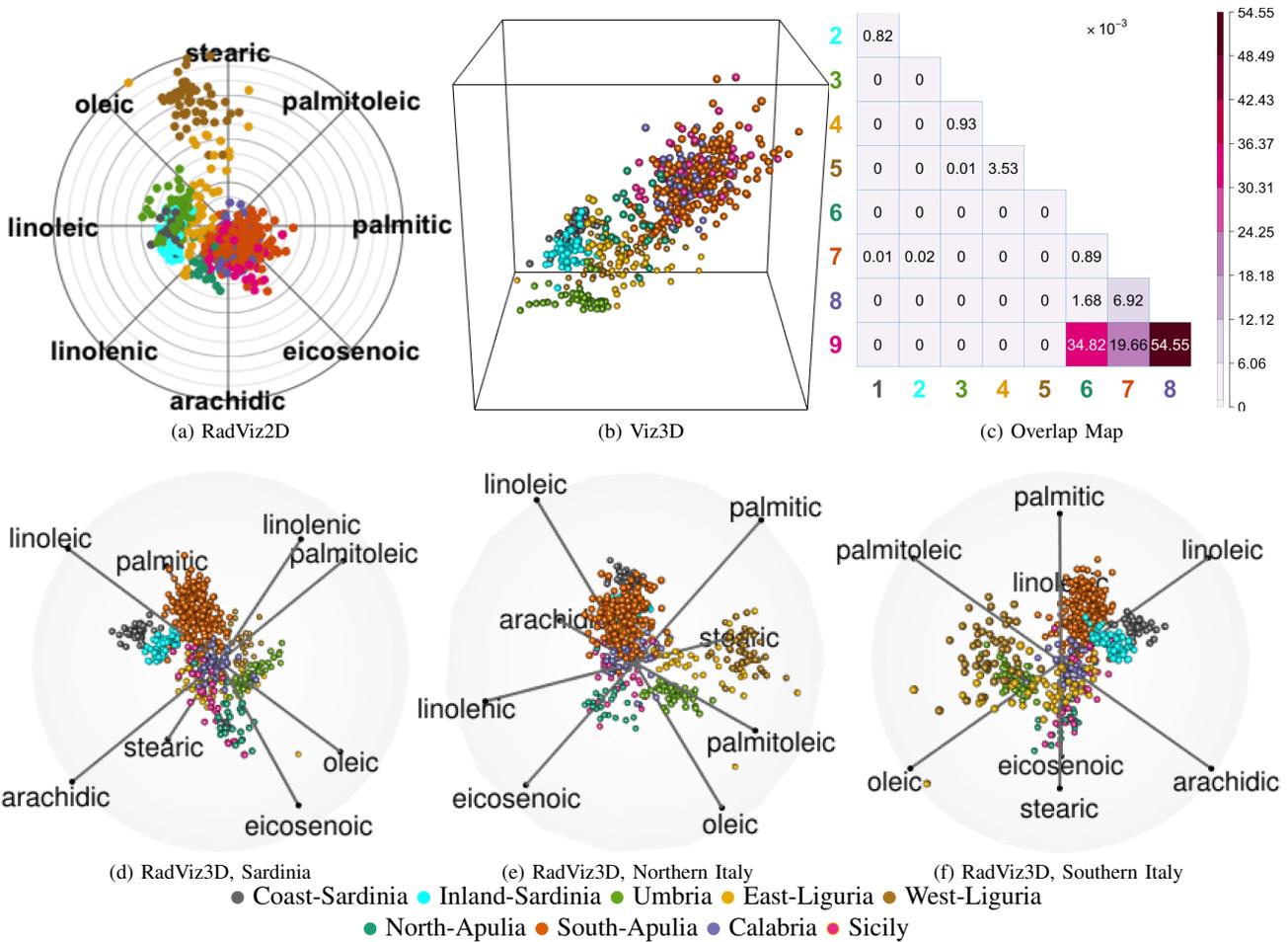
Two regions are (Coastal and Inland) Sardinian, while
three (Umbria, East and West Liguria) are from northern
Italy and four (North and South Apulia, Calabria,
Sicily) are from the South. 
Fig.~\ref{oils} shows West Liguria is separated in all three visualizations,
but  only RadViz3D~(Fig.~\ref{oils}d)  distinguishes coastal and
inland Sardinia oils from the others.  The northern regions are
reasonably distinguished from each other in RadViz3D and well-separated from the
other areas (for both 3D methods, though RadViz3D is clearer). The
southern regions are difficult to distinguish from each  
other but more easily distinguished from the north and Sardinia. 
\citet{almodovarandmaitra18} reported two groups in the southern
oils, with North Apulia more distinct than the rest,
and our displays (Fig.~\ref{oils}d,e) support  this
view. Our displays also echo~\citet{petersonetal17}'s findings that
olive oils are easier distinguished by area than by region.

\subsection{Compositional datasets}
We visualize two compositional datasets not hitherto analyzed in the statistics literature.
\subsubsection{Illustrating the history of Chinese celadon ceramics}
\label{sec:ceramics}
\begin{figure}
\centering
\mbox{
  \subfloat[RadViz2D]{\label{ceramic:radviz2d}\includegraphics[width=0.25\textwidth]{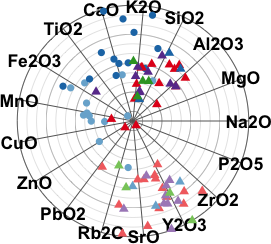}}
  \subfloat[Viz3D]{\label{ceramic:viz3d}\includegraphics[width=0.25\textwidth]{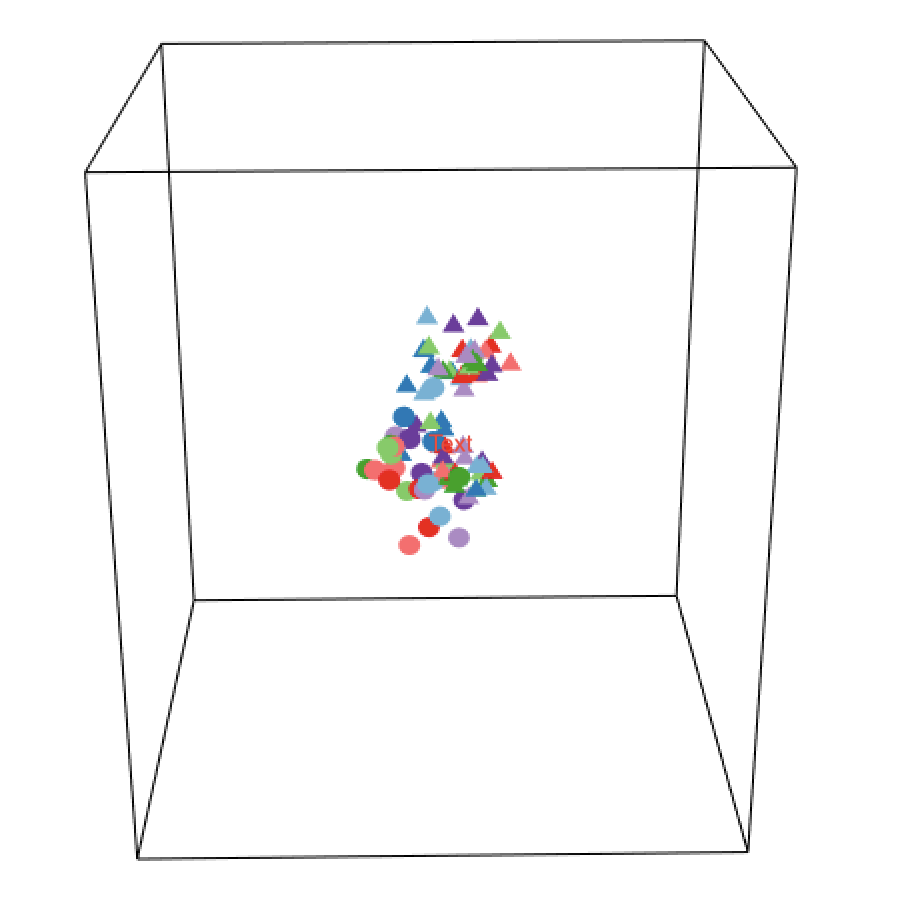}}
  \setcounter{subfigure}{0}
  \subfloat[RadViz3D]{
    \begin{minipage}[b][][t]{.5\textwidth} 
      \mbox{
        \subfloat{\includegraphics[width=0.5\textwidth]{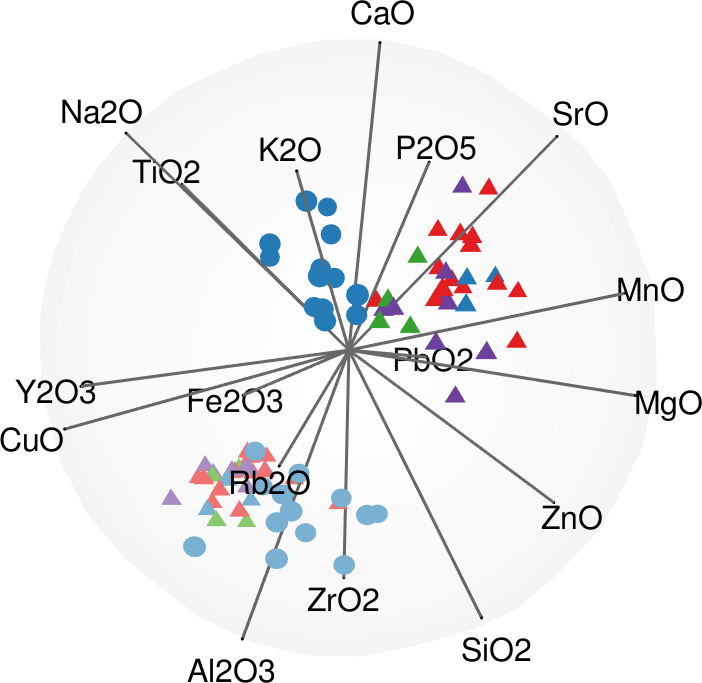}}
        \subfloat{\includegraphics[width=0.5\textwidth]{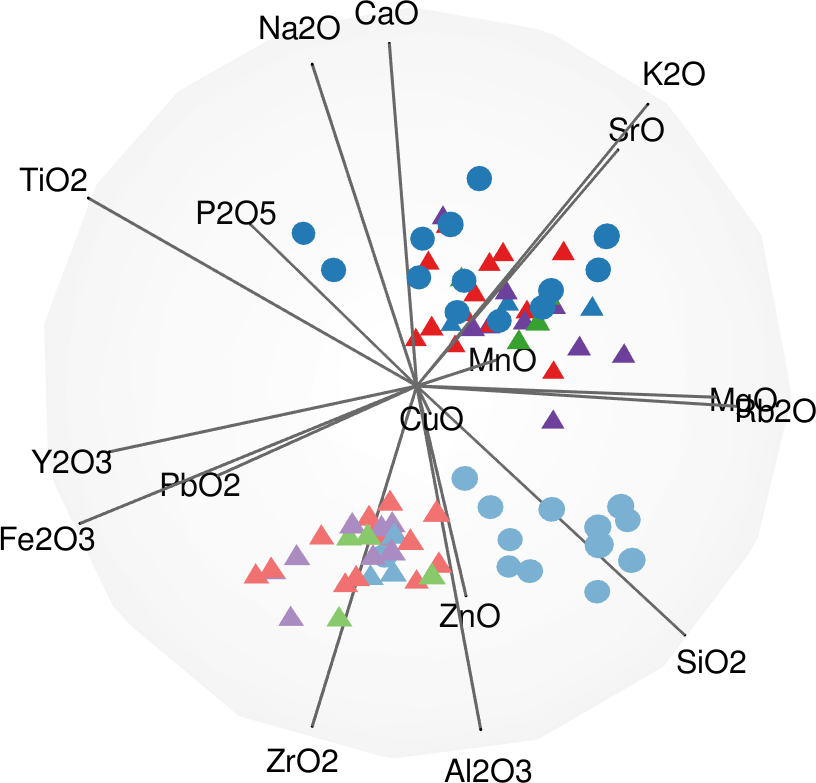}}
      }
    \end{minipage}}
}
\centering
\text{
\tikz\draw[cmp3,fill=cmp3] (0,0) circle (.5ex); Song
\tikz\draw[cmp4,fill=cmp4] (0,0) circle (.5ex); Yuan
\tikz\draw[cmp1,fill=cmp1] (0,0) circle (.5ex); Ming 
\tikz\draw[cmp2,fill=cmp2] (0,0) circle (.5ex); Qing
}\\
\text{
\tikz\draw[cmp6,fill=cmp6] (0,0) circle (.5ex); Body 
\tikz\draw[cmp5,fill=cmp5] (0,0) circle (.5ex); Glaze}
\\
\text{
\tikz\draw[] (0,0) circle (.5ex);  Imitated Longquan
\tikz\draw[]  (0,0)--(0.1,0.2) -- (0.2,0)-- (0,0); Longquan celadon}
\caption{RadViz2D, Viz3D and RadViz3D displays of the {\em celadons}  dataset.}
\vspace{-.1in}
\label{ceramic}
\end{figure}
Longquan celadon~\citep{gompertz80,medley89} is a type of green-glazed Chinese
ceramic with a long history of production at its namesake site. 
It received a major production boost in the Northern Song 
(960--1127) period, continuing under the Southern Song
(1127--1279),  Yuan (1271--1368),  Ming (1368--1644) and Qing
(1644--1912) dynasties. These celadons were much coveted, becoming an 
important part of China's export economy for over five hundred years 
and 
	widely imitated, for instance at Jingdezhen, whose famed blue
and white porcelain finally overtook the Longquan celadon. Samples
of Longquan celadon manufactured in Dayao from the above periods and
their imitations manufactured in Jingdezhen during the Ming period,
were analyzed~\citep{Heetal15} by means of the composition of 17
compounds (Na$_2$O, MgO, Al$_2$O$_3$, SiO$_2$, K$_2$O, CaO, TiO$_2$,
Fe$_2$O$_3$, MnO, CuO, ZnO, PbO$_2$, Rb$_2$O, SrO, Y$_2$O$_3$, ZrO$_2$
and P$_2$O$_5$). In all, chemical compositions of these compounds on
the body and glaze of 44 celadon samples are available, and also
included in the {\tt celadons} dataset of our {\tt radviz3D} package.
Fig.~\ref{ceramic} displays the dataset in terms of the
compositions of the 17 compounds. The RadViz2D and Viz3D displays are
largely uninformative, but RadViz3D illustrates well the history of
Longquan celadon and its imitation. The  chemical composition of
celadon glaze and body are distinct from each other. RadViz3D also
clearly separates Longquan body from its imitation. This accurately
reflects the view that soils used in the manufacture of celadon were
acquired locally by the Jingdezhen craftsmen for reasons of cost due to 
high demand of the celadon body raw materials. Further, archaelogical
excavations~\citep{pengetal09} indicate that Jingdezhen soils have
lower Fe$_2$O$_3$ and  TiO$_2$ content and higher silica (CaO) than
soils at the Dayao sites. As in
\citet{Heetal15}, our display also indicates distinctiveness of
genuine Longquan glaze from its imitation, but there is improvement
over time, so much so that some later Longquan samples
(Qing period) are close in composition to the
Jingdezhen samples. Genuine Longquan chemical composition itself
 evolved over the four dynasties, as indicated in the
figure. In summary, our RadViz3D display of this compositional dataset
provides clearer and more interpretable displays than RadViz2D or Viz3D. 

\subsubsection{Tracking SARS-Cov-2 variant proportions in the US}
\label{sec:application}
The closing months of 2019 saw the emergence of the SARS-Cov-2 virus in
Wuhan, China followed by its rapid spread around the world to become
the most disruptive global pandemic in over a century. 
Several vaccines were developed in record time, 
and multi-phased vaccinations started in several countries,
including in the US in early 2021. However, evolution of the virus
brought new variants, some deadlier and more transmissible.
After the Delta variant arrived in the US, it virtually engulfed
healthcare systems over the summer of 2021. Here, we illustrate the
changing proportions of the SARS-Cov-2 variants in the ten
US HHS regions during the rapid transition to Delta.

Our dataset is from the most recent (per October 5, 2021) weekly
average proportions of 12 sets of variants (recorded from June 5,
2021 and up to the week of September 18, 2021) made publicly
available as part of the National SARS-Cov-2 Genomic Surveillance
Program of the Centers for Disease Control and Prevention (CDC). 
The time period starts just $10$ days before the CDC declared Delta a Variant Of Concern in the US because of its increased transmissibility and potential reduction in antibody neutralization. 
We calculated the
running three-week average proportions of each variant (with averages
computed after transforming each proportion by the arc sine
of its square root, and transforming back after calculating the
three-week average). These three-week average variant proportions for
each of the ten HHS regions are in the {\tt sarscov2.us.variants} dataset in the
{\tt radviz3d} package and displayed via RadViz3D in Fig.~\ref{covid}
and RadViz2D and Viz3D (see supplement). 
\begin{figure}[h]
  \centering
  \begin{minipage}[t]{0.62\textwidth}
      \subfloat{\label{covid}\includegraphics[width=\textwidth]{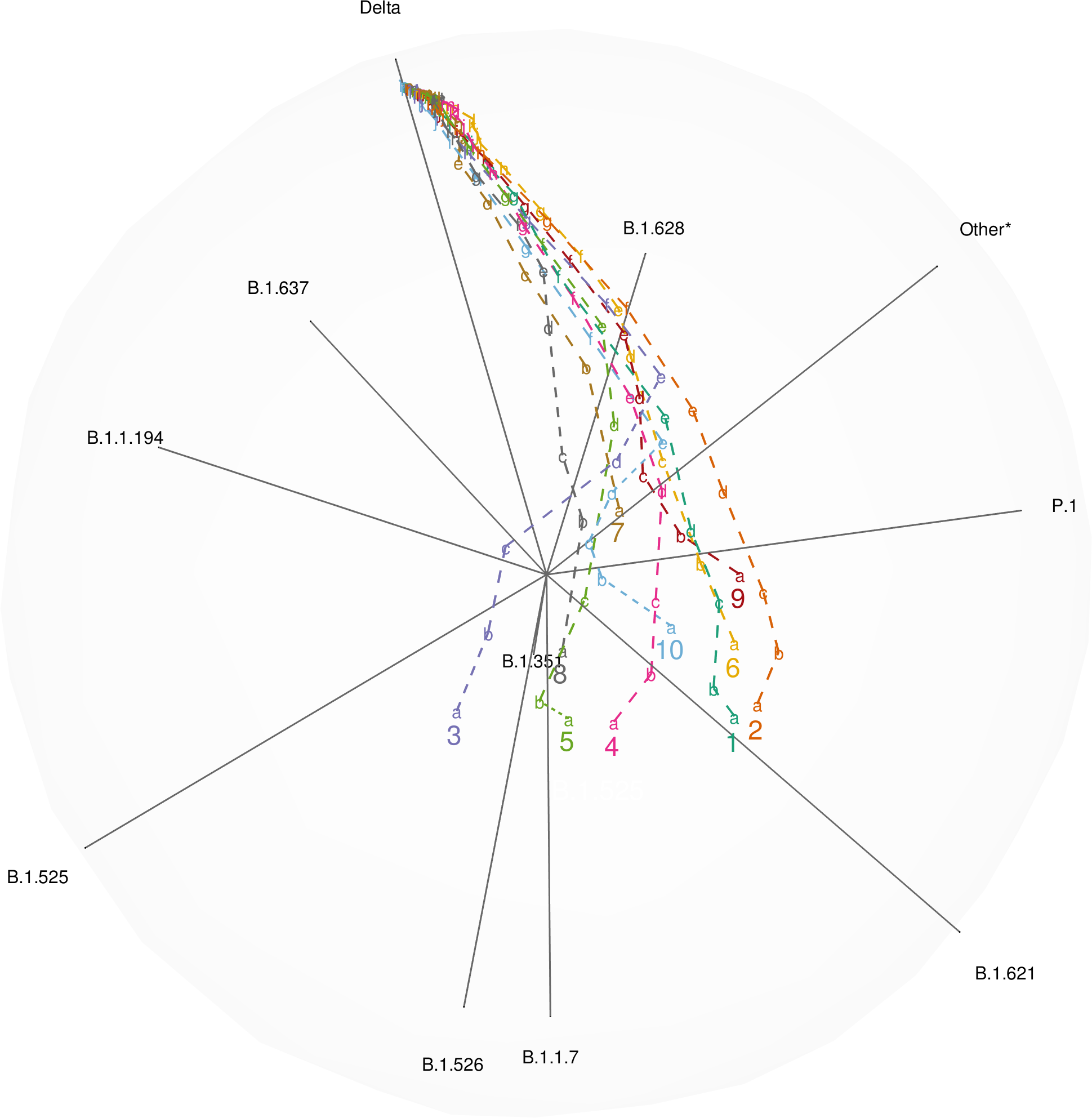}}
    \end{minipage}%
    \begin{minipage}[t]{0.38\textwidth}
      \subfloat{\resizebox{\textwidth}{!}{
        \begin{tabular}{l|l}
          \hline
          \bf{Variant Lineage} & \bf{WHO Label} \\ \hline
          B.1.1.7 & Alpha\\
          B.1.351 & Beta\\
          P.1 & Gamma\\
          B.1.617.2, AY.1, AY.2 & Delta\\
          B.1.525 & Eta\\
          B.1.526 & Iota\\
          B.1.617.1 & Kappa\\
          B.1.621 & Mu\\
          B.1.628, B.1.637, B.1.1.194 & Unassigned\\
          \hline 
        \end{tabular}
      }}\par
    \subfloat{\includegraphics[width=\textwidth]{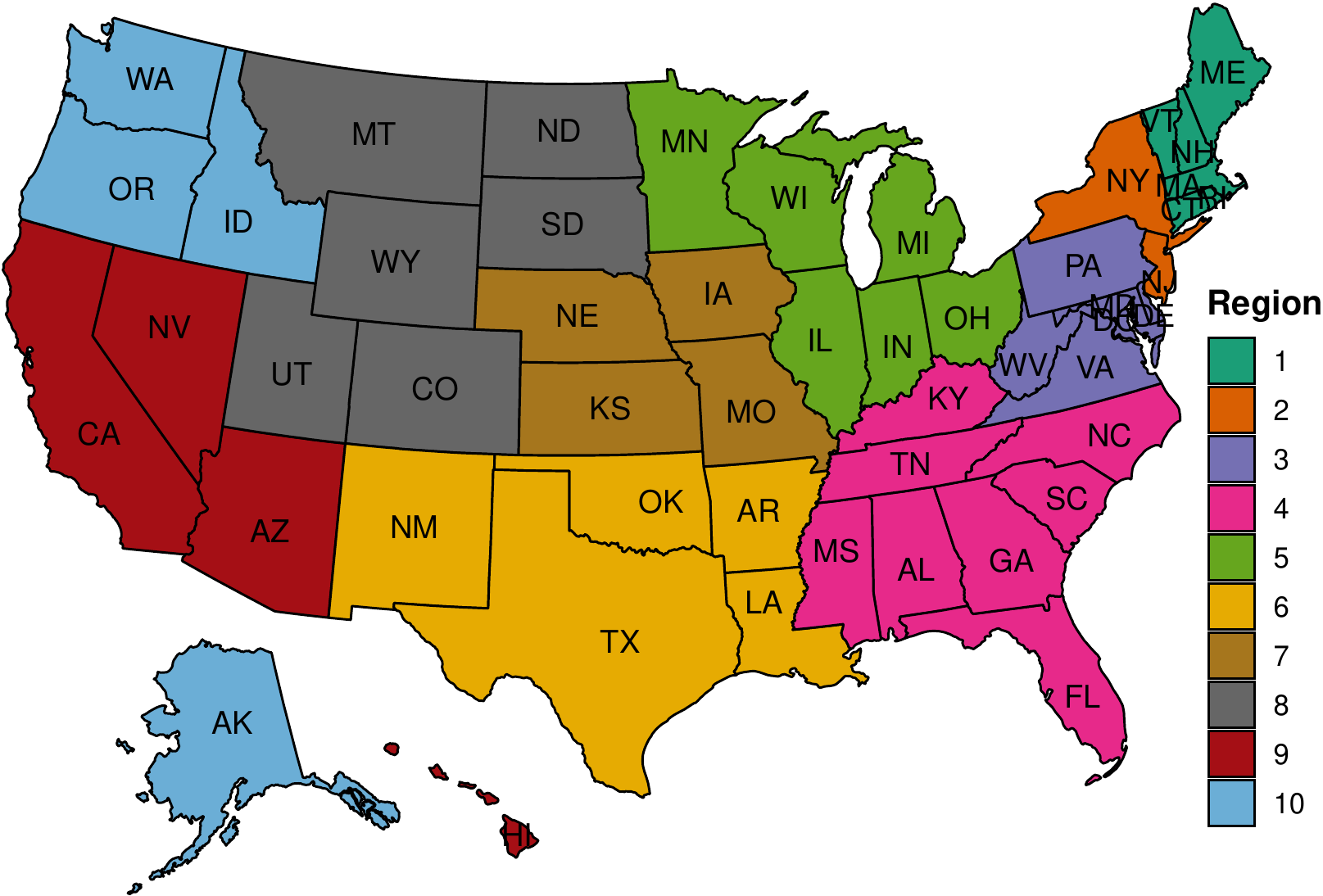}}
    \\
    {\small United States HHS Regions. (Note: Region 2        includes
    Puerto Rico and the Virgin Islands, while Region 9        includes
    American Samoa, Micronesia, Guam, Marshall Islands, Northern
    Marianas Islands and Palau.)}
    \end{minipage}
      \caption{RadViz3D display of the SARS-Cov2 variant proportions
        (top right corner)
        over the summer of 2021 in the ten HHS regions of the US
        (indexed by map on bottom  right corner). Half-broken lines
        connect the average variant proportions over the three weeks ending on
        (a) June 19,  (b) June 26, (c) July 3, (d) July 10, (e) July 17, 
        (f) July 24, (g) July 31, (h) August 7, (i) August 14, (j)
        August 21, (k) August 28, (l) September 4, (m) September 11 and (n)
        September 18.      }      \label{covid}
  \end{figure}
The display demonstrates how the Delta variant rapidly eclipsed 
all SARS-Cov-2 infections across all ten HHS regions by the end of the time period.
Our display shows the Delta variant with a notable head start in Region 7 where it was already the majority strain (65.9\%) in the first time period under consideration.
Delta also started with a weak majority (at 51.2\%) in Region 8. 
The Alpha variant dominated most other regions in early June 2021; it is the majority variant in Regions 3--6 and 10.
There was much greater diversity of strains during the dominance of Alpha in early June (time point (a) in Fig.~\ref{covid}) compared to mid-September (time point (n)) during the dominance of Delta.
Entropy varied from 0.90 in Region 7 to 1.67 in Region 1 in early June (0.75--1.49, excluding Delta), but decreased to a maximum across regions of 0.04 in Region 6 in mid-September.
Despite the three-week averaging, there are fluctuations in variant compositions among time periods, especially in Regions 10 and (to a lesser extent) 3.
The rapid fluctuations may be sampling error, but it is also evident that several variants were on the rise before Delta swept through.
While Alpha monotonically declined, the next overall most prevalent variant Gamma 
	peaked in weeks 2--3 in the northeast (Regions 1--3) and Region 8.
The next most prevalent variant Mu (B.1.621) peaked east of the Mississippi (Regions 1--5) in weeks 2 and 3, in the west (Regions 9 and 10) in week 4 or 5, and displayed mixed behavior in the midwest (Regions 6--8).
Other notable increases include a brief appearance of B.1.637 during weeks 2--4 in all regions but 7,
	B.1.628 increasing in all regions until about week 4 (July 10), with a slight two-week delay in Regions 7 and 8, 
	and B.1.1.194 rising in prevalence in regions 3, 4, 9 and 10 in early June. 
These trends in the lesser variants likely explain why the trajectories do not simply traverse a straight line from Alpha to Delta, but veer toward P.1 (Gamma) or Mu (B.1.621), with nuances dictated by the mix of variants present in the HHS region at the start of the transition.
We also caution that some of the patterns may be due to biases in variant sampling. 
It is plausible that patients with more severe disease, possibly Delta in particular, may seek medical care, increasing the chance of Delta inclusion in the surveillance sample, especially perhaps in regions with low testing rates.
Regardless of the cause of the observed trends, RadViz3D very
effectively illustrates the changing composition of the SARS-Cov-2
virus variants over the course of the summer of 2021.

\section{Discussion}
\label{sec:conclusion}
This short technical note 
shows that placement of anchor points in a radial visualization plot
should  ideally be as far away from each other as possible to avoid
artificially-induced  visual correlations between the
coordinates. This desired property motivates 
our development of a fully 3D radial visualization tool called
RadViz3D as it provides for larger  
distances between anchor points. Our development hinges on our exact and
approximate solutions for the placement of $p$ anchor points on the 3D
sphere.
A R package {\tt radviz3d} 
implementing our methodology is publicly available at
\url{https://radviz3d.github.io}, and is demonstrated to
accurately reveal class distinctions and lower-dimensional structure
in simulated and standard classification datasets. As pointed out by a
reviewer, the radial visualization methodology is naturally suited to
displaying compositional data, so we have used RadViz3D to display to
illustrate the historical development, by means of its chemical
composition, of Longquan celadon and its Jingdezhen imitation. We also used the
methodology to illustrate the changing face of the SARS-Cov-2 pandemic
over the US regions during the summer of 2021. 

Some aspects of our development could benefit from further
attention. For instance, our development of RadViz3D
demonstrates that we should use (at least approximately) equi-spaced
anchor points, but 
the order of the anchor points may still be material. 
The order of anchor points corresponds to switching the order of the
columns in the projection matrix $\bm U$ in~\eqref{eq:T}. Clearly
different orders of anchor points will produce different
visualizations. For the visualization of grouped data, one possible
solution to this is to run through all possible orders and pick one
that has the biggest separation between groups, where $\ddot\omega$
can be used to measure separation. However, this would be
very computationally demanding in higher dimensions. Also,
based on our assumption that all coordinates 
are uncorrelated with equi-spaced anchor points, the order is intuitively perhaps not as important
since all the coordinates are somewhat equivalent in our display. But
it would still be worth getting a definitive answer. Further, for
high-dimensional datasets, we will need to look at reducing
dimensionality in data before using RadViz3D. \citet{laaetal21}
recently framed the grand
tour~\citep{asimov85,bujaetal05} in the context of radial
visualization, with improvements for the ``reverse curse of
dimensionality''~\citep{diaconisandfreedman84,marronetal07,ahnandmarron10}
or address data clumping or piling seen in the case of such displays
with high-dimensional data, so it  may be worth incorporating their
ideas to potentially  improve RadViz3D. Thus, we see that
while this paper has made possible fully 3D radial visualization,  there remain
issues that merit additional investigation.



%



 \section*{Acknowledgments}
The authors thank three anonymous reviewers and an anonymous Associate
Editor for their insightful comments on an earlier version of the
manuscript. We are also grateful to N. Kunwar, H. Nguyen, P. Lu,
  F.S. Aguilar, G. Agadilov and I. Agbemafle for helpful
  discussions during an introductory graduate multivariate
  statistics class (Stat 501, Spring 2018 semester) at Iowa State University where this
  methodology was   conceived. Our thanks also to Z. He and H. Zhang for their
  explanations on how the {\tt ceramics} observations were collected
  and to K. S. Dorman for her help in identifying and obtaining the
  SARS-Cov-2 variants proportions dataset. 
 The third author's  research was
supported in  part by the United States Department
  of Agriculture (USDA)/National Institute of Food and
  Agriculture (NIFA), Hatch projects IOW03617 and IOW03717. The content of this paper however is
  solely the responsibility of the  authors and does not represent the
  official views of the USDA. 
\bibliographystyle{IEEEtran}
\bibliography{references}



%



\end{document}